\documentclass[onecolumn,epsfig]{elsart}
\usepackage{epsfig}

\usepackage{psfig}
\usepackage{epsfig}
\usepackage{floatflt}
\usepackage{graphicx}
\usepackage{dcolumn}
\usepackage{bm}

\newcommand{\AmS}{{\protect\the\textfont2A\kern-.1667em\lower.5ex\hbox{M}\kern-.125emS}}
\def\Journal#1#2#3#4{{#1} {\bf #2}, #3 (#4)}

\def\NPB{{ Nucl. Phys.} B}
\def\NPA{{ Nucl. Phys.} A}
\def\PLB{{ Phys. Lett.}  B}
\def\PRL{ Phys. Rev. Lett.}
\def\PR{ Phys. Rep.}
\def\PRC{{ Phys. Rev.} C}
\def\PRD{{ Phys. Rev.} D}
\def\ZPC{{ Z. Phys.} C}

\def\EPJC{{ Eur. Phys. J.} C}
\def\NJP{New J. Phys.}

\def\be{\begin{equation}}
\def\ee{\end{equation}}
\def\bea{\begin{eqnarray}}
\def\eea{\end{eqnarray}}

\begin{document}
\begin{frontmatter}
\title{Universal anti-baryon density in $e^+e^-$,$\gamma p$, pp, pA and AA collisions}

\author[ustc,bnl,lbl]{Haidong~Liu}, \author[bnl]{Zhangbu~Xu}
\address[ustc]{University of Science \& Technology of China, Anhui 230027, China}
\address[bnl]{Brookhaven National Laboratory, Upton, New York 11973}
\address[lbl]{Lawrence Berkeley National Laboratory,
Berkeley, California 94720}

\date{\today}%
\begin{abstract}
We compiled the systematical measurements of anti-nucleus production
in ultra-relativistic heavy ion collisions as well as those in $pp$,
$p\bar{p}$, $\gamma p$ and $e^{+}e^{-}$ at various beam energies.  The
anti-baryon phase space density inferred from $\bar{d}/\bar{p}$ ratio
in $A+A$, $p+A$, $pp(\bar{p})$ and $\gamma p$ collisions is found to  
follow a universal distribution as a function of center of mass of beam  
energy and can be described in a statistical model. We demonstrated  
that anti-baryon density in all the collisions is the highest when the  
collisions are dominated by the processes of $g+g$ or $\bar{q}+g$. In
$e^+e^-$ collisions at LEP, the cross section of $q\bar{q}g$ is  
suppressed by a factor of strong coupling constant $\alpha_s$ relative  
to $q\bar{q}$. This can consistently explain the $\bar{d}$ suppression  
observed by ALEPH relative to that in $e^+e^-\rightarrow ggg$ by  
ARGUS. We discuss the implications to the baryon enhancement at high  
transverse momentum at RHIC when jet is quenched.

\end{abstract}
\end{frontmatter}

\section{Introduction}
    Relativistic heavy ion collisions create high energy density
and high baryon density in the reaction zone. Light nuclei and
their antiparticles can be produced by the recombination of created
nucleons and anti-nucleons or stopped
nucleons~\cite{nagle,heinzcoalescence,llope:95,coaltheory}. This recombination
process is called coalescence. Since the binding energy of nucleus is
small, coalescence can only happen at the late stage of the evolution
when the interactions between nucleons and other particles such as
pions are weak. Therefore, the production of nucleus provides a tool
to measure baryon distribution at the thermal freeze-out where the
interactions between particles are weakening. Since the probability of
coalescence of a particular nuclear system (d, $^{3}He$, etc.) depends
on the properties of the hadronic system formed at late stage as a
result of the collision, its evolution and hadronization, the study of
the coalescence process is useful in elucidating those properties. For
example, in a coalescence model, the coalescence probability depends
on the temperature, baryon chemical potential (essentially the baryon
density), and the size of the system, as well as the statistical
weight (degeneracy) of the coalesced nucleus~\cite{E864A}. From the
measurement of the nucleus production, we will be able to construct
the thermal freeze-out in the ($T$, $\mu_B$) phase
diagram~\cite{heinzcoalescence}. Together with the measurements of
other particle yields from which the statistical model can construct
the chemical freeze-out, we will be able to have a better
understanding of how the system evolves from chemical to thermal
freeze-out.

How baryons are created and interact among themselves and with other
particles in the course of the evolution of the system is an important
subject in relativistical heavy ion collisions. The coalescence
process is not unique in relativistic heavy ion collisions. In fact,
the idea was originated from elementary particle collisions where it
was used to formulate nucleus
production~\cite{earlypp}. Systematically measuring the coalescence
effect of nucleons in collisions such as $e^{+}e^{-}$, $pp$,
$\bar{p}p$, $pA$ and $AA$ may give us insight in this
subject. Analytic formulae from coalescence
models~\cite{nagle,heinzcoalescence,llope:95,fqwang,coaltheory} can be
used to extract the information. Transport
models~\cite{nagle,coaltheory} with detailed phase space distribution
of nucleons at coalescence have been used as well.

The formula of coalescence is of the form:
\begin{equation}
E_{A}{{d^{3}N_{A}}\over{d^{3}p_{A}}} =
B_{A}(E_{p}{{d^{3}N_{p}}\over{d^{3}p_{p}}})^{Z}
(E_{n}{{d^{3}N_{n}}\over{d^{3}p_{n}}})^{A-Z}
\label{eq:ba}
\end{equation}
where $E{{d^{3}N}\over{d^{3}p}}$ is the invariant yield of nucleons or
nuclei, A is the nuclear number of the produced nucleus and N, Z are
the numbers of neutron and proton in the nucleus respectively.  the
differential cross section of the anti-deuteron is denoted as
$\bar{d}$ and that of the anti-proton as $\bar{p}$ for simplicity in
some cases in the following discussions. Simple coalescence, density
matrix, sudden approximation and thermodynamic models predict a
slightly different expression of $B_{A}$~\cite{nagle}. The fragment
coalescence model\cite{llope:95} can be used to extract the source
radii for different nuclei at their 'freeze-out' from its yields and
the yields of smaller fragments. If assuming Gaussian sources, for the
case of A=2 (deuteron) from nucleon coalescence (ignoring isospin
difference):
\begin{equation}
\label{eq:phase}
{\frac{d^{3}N_{d}}{d^{3}p_{d}}}=
{\frac{\pi^{3/2}{\frac{d^{3}N_{n}}{d^{3}p_{n}}}{\frac{d^{3}N_{p}}{d^{3}p_{p}}}}
{{(R_{G}^{2}+\rho_{d}^{2}/2)}^{3/2}}}\exp{(\Delta E/T)}
\label{eq:rg}
\end{equation}
where $\rho_{d}$ is the size parameter of the composite's wave
function~\cite{llope:95}, $p+n\rightarrow d+\Delta E$, $\Delta E$ is
the excess of binding energy for that channel.  This source radius
($R_{G}$) should be comparable to the source size from two-particle
interferometry (HBT). In collisions with small reaction volume such as
in pp and pA, the size of the nucleus and a maximal relative momentum
($p_{0}$) between the coalescing nucleons dominate
($\rho_{d}>>R_{G}$). This results in an approximate constancy of
$B_{A}$ parameter characterized as momentum space coalescence volume,
independent of beam energies and beam species. On the other hand, the
phase space density is $[4/3/(2\pi)^2]{\bar{d}/\bar{p}}$ regardless of
the system~\cite{fqwang}. Recent development began to focus on the
phase space aspects of the coalescence process which also tried to
take into account the dynamical expansion.  For the coalescence with
nucleons and nuclei of low transverse momentum and middle rapidity,
Eq.2.14b in Ref.~\cite{heinzcoalescence} can be further simplified as:
\begin{equation}
{\frac{d^2N_{A}}{2\pi{p_{T}}dp_{T}dy}}\simeq{\frac{2J_{A}+1}{(2\pi)^{3}}}e^{({\frac{\mu_{A}-M}{T}}
)}
M_{T}V_{eff}e^{(-\frac{M_T-M}{T^{*}})}
\label{eq3}
\end{equation}
The ratio of the differential cross sections of anti-deuteron (denoted
as $\bar{d}$) to anti-proton (denoated as $\bar{p}$) is therefore
related to the phase space density of the nucleon~\cite{fqwang}, the
baryon chemical potential and thermal freeze-out temperature in
coalescence model.  In this paper, we will mainly discuss the
experimental results in terms of phase space density, baryon chemical
potential and thermal freeze-out temperature from coalescence
measurements.  Most of the discussion and calculation are related to
Eq.~\ref{eq3}.  The quantity of
${\frac{2J_{A}+1}{(2\pi)^{3}}}M_{T}V_{eff}e^{(-\frac{M_T-M}{T^{*}})}$
depends on details of the dynamics of the system, quantum effect,
binding energy of nucleus and the selected $p_{T}$ range.  This is
taken as a constant ($C$) for coalescence at low $p_{T}$ for a given
colliding system
\begin{equation}
{\frac{d^2N_{A}}{2\pi{p_{T}}dp_{T}dy}}\simeq{C}e^{(-{\frac{(m_{B}\pm\mu_{B})A}{T}})}
\label{eq4}
\end{equation}

\section{Chemical and Thermal Properties of Anti-nucleus Production}

\begin{figure}
  {\includegraphics*[width=2.7in]{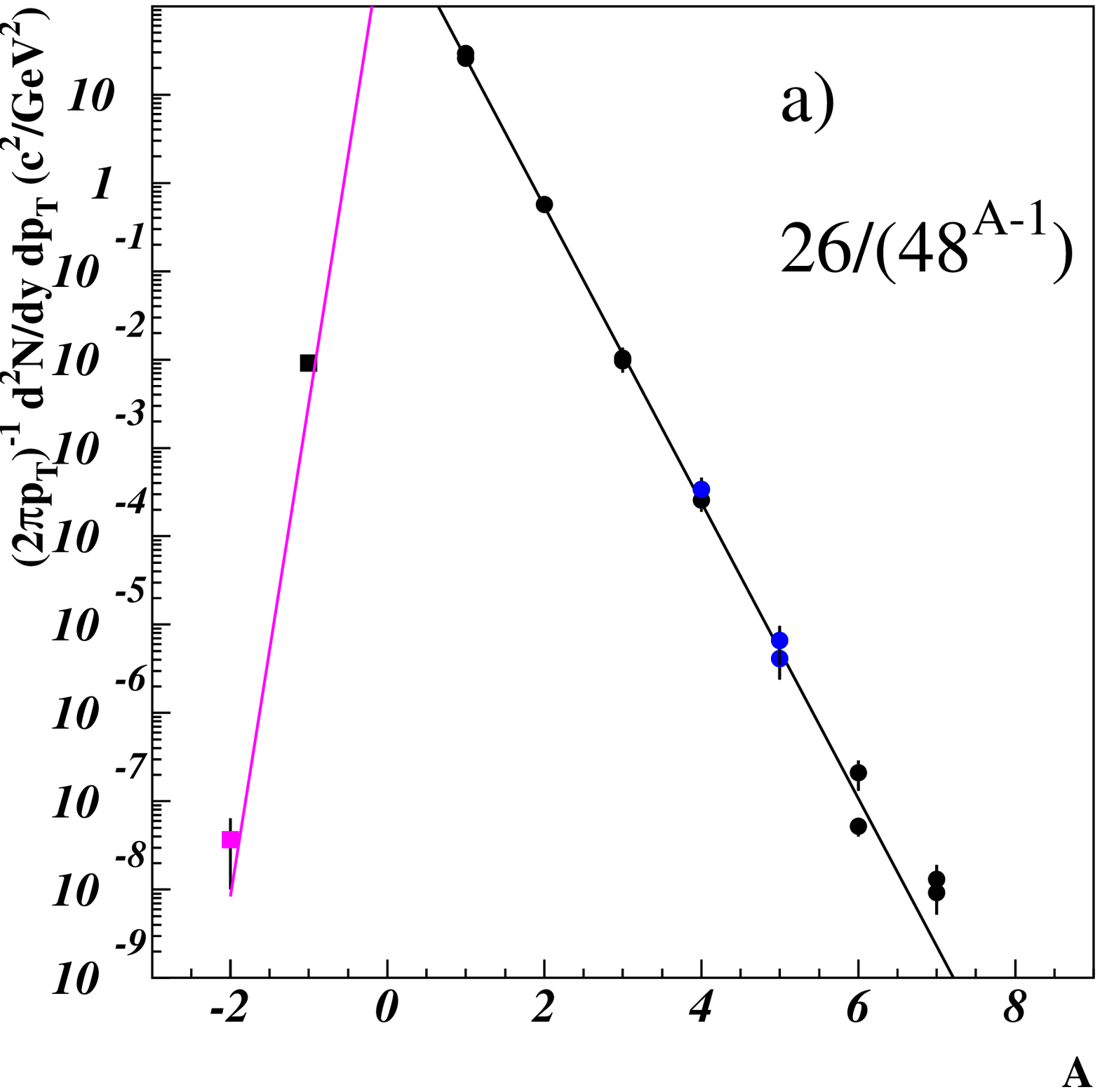}}
  {\includegraphics*[width=2.7in]{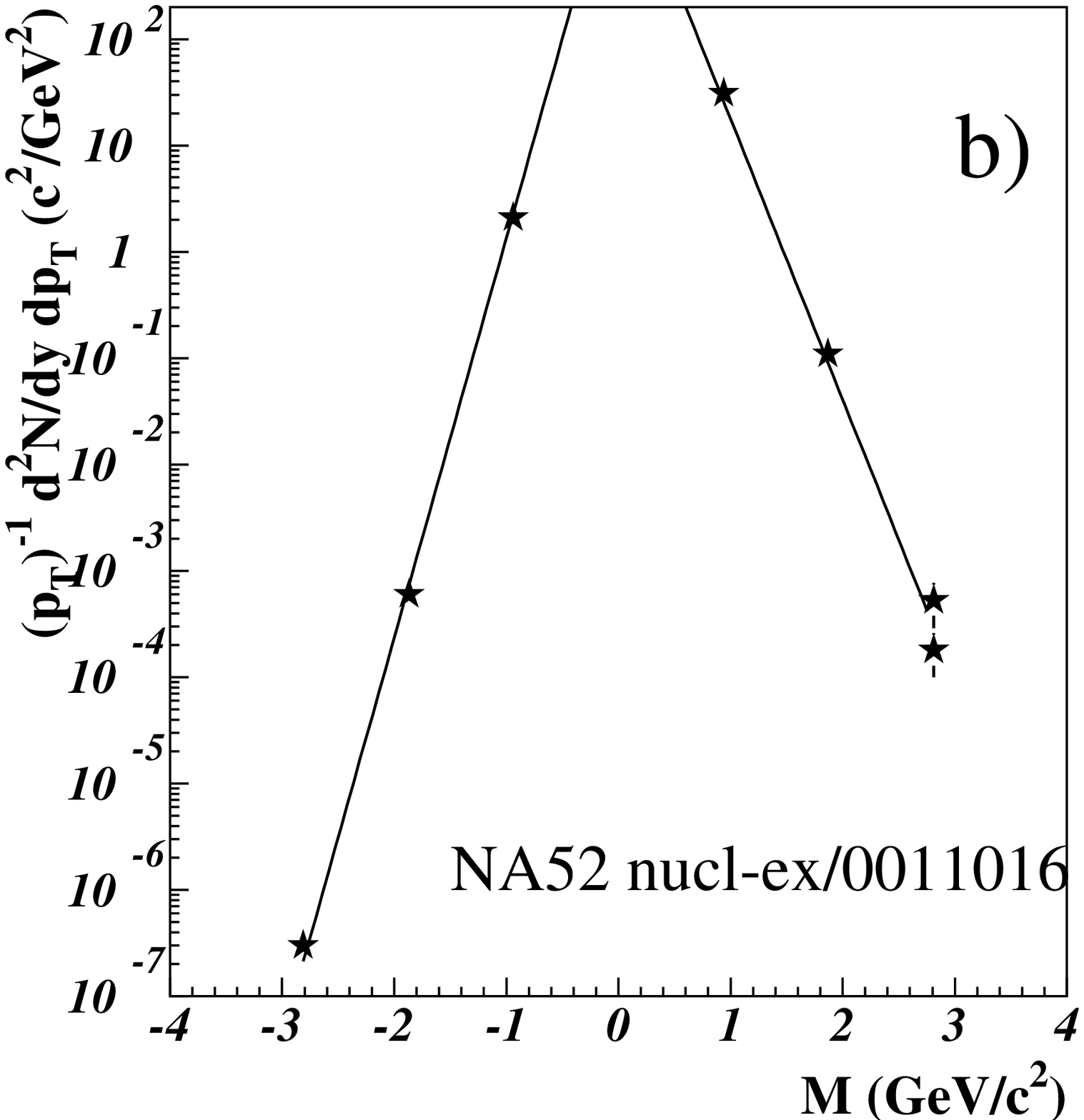}}
  {\includegraphics*[width=2.8in]{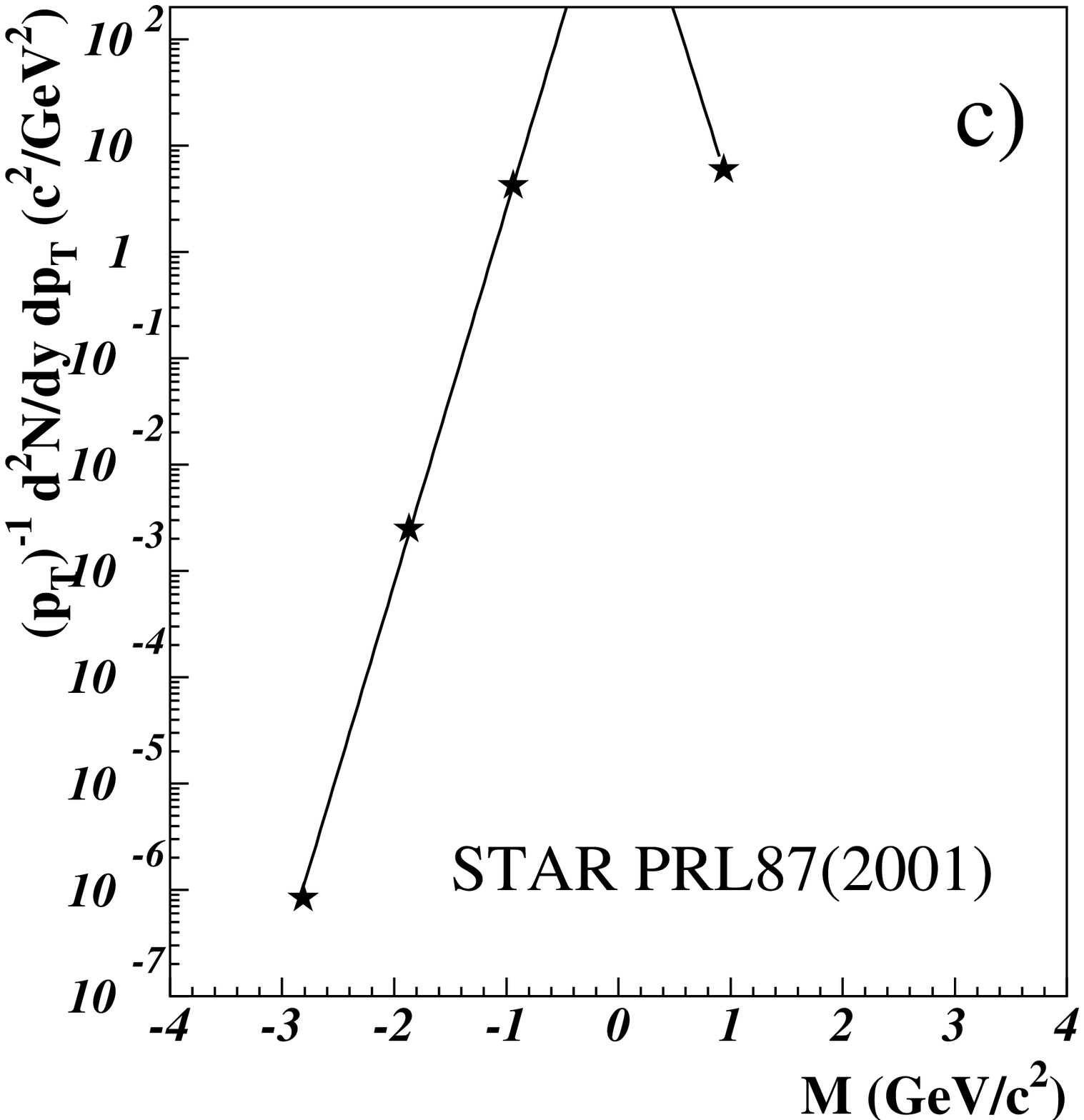}}
  {\includegraphics*[width=2.8in]{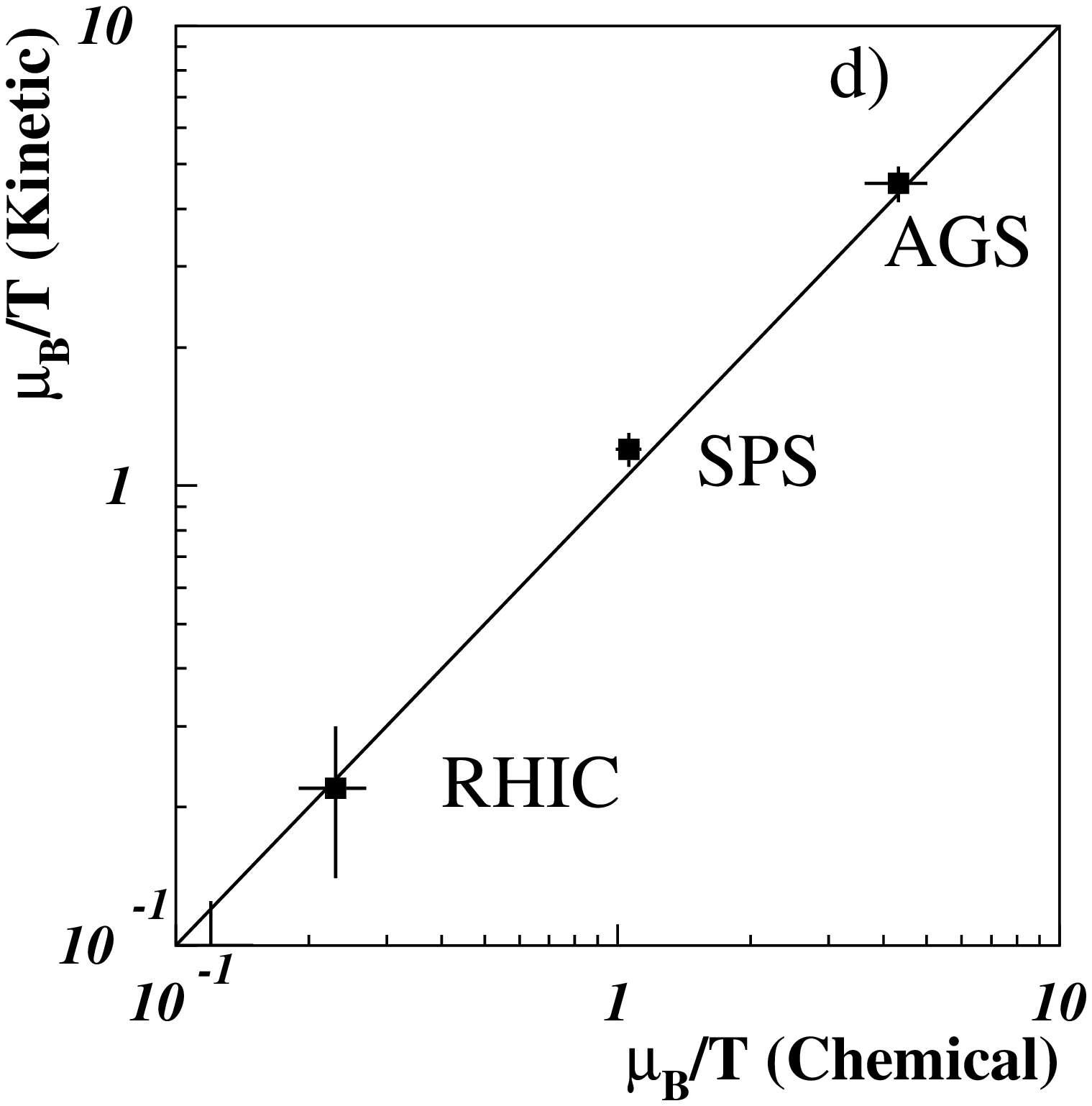}}
  \caption{a) Eighteen hadrons measured by E864~\cite{E864}. Stable
light nuclei\protect\cite{e864xzb} are selected from \(y=1.9\) and
\(p_{T}/A<300MeV/c\). ${\bar p}$ is at $y=1.9$ and
$p_{T}\simeq0$\protect\cite{pbar}. ${\bar d}$ is at $y=1.9$ and
$0<p_{T}<1GeV/c$. Lines are fits to the data assuming local thermal
equilibrium. See text for details. Similar measurements are from
NA52/SPS~\cite{NA52} and STAR/RHIC~\cite{stardbar} and PHENIX/RHIC at
higher $p_T$~\cite{phenixdbar} as presented in b) and c),
respectively. d)The measured $\mu_{B}/T$ at thermal freeze-out vs that
at chemical freeze-out.
\label{fig:all18hadrons}}
\end{figure}

  Production of light nuclei in heavy ion collisions has been measured
by many groups at different beam energies and colliding
systems\cite{nigel:thesis} from BELAVAC to RHIC energies.  A
reasonable list of the measurements can be found in
\cite{nigel:thesis}. The results discussed here are results from E864
at the AGS~\cite{E864,E864A,E864Hes}, NA52 at SPS~\cite{NA52} and STAR
at RHIC~\cite{stardbar}.  Fig.~\ref{fig:all18hadrons} shows the
invariant yields of 18 hadrons being measured in a selected rapidity
($y$) and transverse momentum ($p_{T}$) bin by E864. It has been
observed that a striking exponential dependence of light nuclei
production up to $A=7$ over ten orders of magnitudes in the invariant
yield including nuclei and
anti-nuclei~\cite{E864A,E864,E864Hes}. Invariant yields are calculated
and presented in terms of $d^{2}N/(2\pi p_{T}dp_{T}dy)$ in units of
$GeV^{-2}c^{2}$ per central collision.  E864 Collaboration has fitted
the production of the ten stable light nuclei with an exponential as a
function of nuclear (baryon) number. The best fit\cite{e864xzb} is
$26/48^{A-1}$.  

The difference between fugacity of proton $\lambda$
and neutron $\lambda^{\prime}$ comes from the difference of the
isospin abundance or $n/p$ ratio in a thermal distribution. From
neutron measurement\cite{evan} of $n/p=1.19\pm0.08$, we have
$\lambda^{\prime}=1.1\lambda$.  For simplicity, the average fugacity
or chemical potential are taken as:
$\lambda_{B}=\sqrt{\lambda\lambda^{\prime}}$, and $\mu_{B} =
(\mu_{p}+\mu_{n})/2$.
Due to secondary nucleus production from beam pipe and limited $p_{T}$
range of PID from $dE/dx$~\cite{stardbar}, STAR is only able to
identify and measure light anti-nuclei ($\bar{d},\overline{^{3}He}$),
$\bar{p}$ and proton. Fig.~\ref{fig:all18hadrons} (right panel) shows
the measured differential yield at $p_{T}/A\simeq0.4$ GeV/$c$ as
function of particle's atomic number $A$ for
$p,\bar{p},\bar{d},\overline{^{3}He}$.

Using Eq.~\ref{eq4}, we are able to fit well the nucleus production at
low $p_{T}$ at AGS, SPS and RHIC heavy ion collisions and obtain
temperature and baryon chemical potential at kinetic freeze-out. The
different slopes of baryons and antibaryons allow the determination of
$\mu_B$. The results are: $T=126MeV$, $\mu_{B}=21MeV$ (RHIC);
$T=130MeV$, $\mu_{B}=170MeV$ (SPS); $T=110MeV$, $\mu_{B}=500MeV$
(AGS). Errors were estimated to be $\pm10MeV$ for all the
cases~\cite{e864xzb}.  Fig.~\ref{fig:all18hadrons}.d shows the ratios
of $\mu_{B}/T$ at kinetic freeze-out vs that at chemical freeze-out at
AGS, SPS and RHIC central AA collisions. The measurements are indeed
consistent with each other even though the parameters at chemical
freeze-out are extracted from thermal fit to the many ratios of the
integrated particle yields~\cite{chemicalfreezeout} while the
parameters from coalescence are from nuclear cluster formation rate at
low $p_{T}$.  However, whether such a trajectory along the phase
diagram is possible requires further study since some of the
studies~\cite{rapp} suggest a path of constant ratio of entropy to
baryon number where as temperature drops baryon chemical potential
increases.  The coalescence measurement suggests a path corresponding
to large entropy increase from chemical to thermal freeze-out.

\section{Anti-baryon Density}

\begin{figure}
  {\includegraphics*[width=5.in]{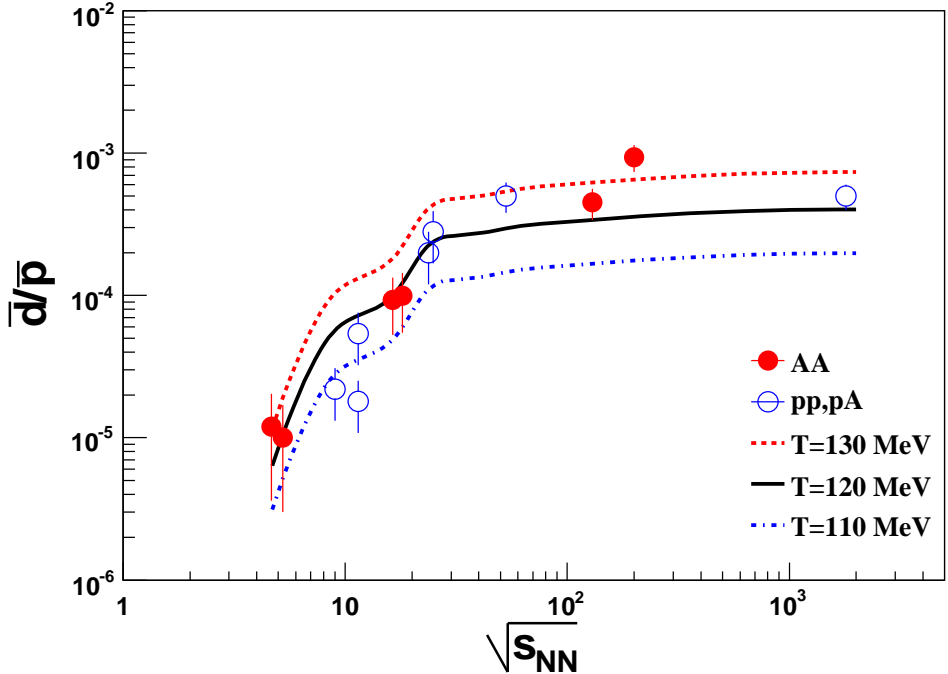}}
  \caption{$\bar{d}/\bar{p}$ as a measure of antibaryon phase space
density as a function of beam energy for pp, pA and AA collisions. The
curves are $\exp{(-m_{B}/T)}\times\sqrt{\bar{p}/p}$ for three choices
of T at 130MeV, 120MeV and 110MeV.} \label{fixt}
\end{figure}

The coalescence measurement is also sensitive to the phase space of
baryon and anti-baryon~\cite{fqwang}.  For example, $\bar{d}/\bar{p}$
(ratio of the differential cross sections) can be taken as a measure
of the anti-baryon phase space density at kinetic freeze-out where
coalescence happens.  Data of $\bar{d}/\bar{p}$ ratio from various
beam energies and colliding species
($pp,\bar{p}p,pA,AA$)~\cite{ISRpA1,ISRpA2,ISRpp,E735,E858,E864A,E864,NA52,NA44,stardbar,phenixdbar}
were collected and shown in Fig.~\ref{fixt}. One very interesting
observation is that the ratio increases monotonically with beam
energies and reaches a plateau above ISR beam energy regardless of the
beam species ($pp,pA,AA$).  Similar behavior has been seen in
$\bar{p}/p$ ratio as a function of beam energy~\cite{pbarp}. This
similar trend is quite nature in terms of thermal model and
coalescence as shown in Eq.~\ref{eq3}.  The relation between
$\bar{d}/\bar{p}$ and $\bar{p}/p$ is
$$\bar{d}/\bar{p}\simeq\exp{(-m_{B}/T)}\times\sqrt{\bar{p}/p}$$ where
$m_{B}$ is the nucleon mass and $T$ is the freeze-out temperature. The
curves in Fig.~\ref{fixt} correspond to three choices of
$T=130,120,110$ MeV. This relation shows qualitative agreement with a
fixed freeze-out temperature at about 120MeV. Recently, CERES
Collaboration observed a universal thermal freeze-out
behavior~\cite{ceres}. These two measurements may be closely
related. However, It also indicates that the temperature may slightly
depend on beam energy with lower temperature at lower $\sqrt{s}$ and
higher value at higher $\sqrt{s}$. These findings suggest that the
anti-nucleons are produced and coalesce in a statistical fashion in
$A+A$, $p+A$, $p+p$ and $\gamma p$ collisions at similar density.
Final state interactions in nucleus-nucleus collisions do not change
this aspect. It also shows that correlations among anti-nucleons in
momentum and coordinate space do not alter the $\bar{d}$ yields since
its production can be described statistically.

\section{Microscopic Processes Creating Anti-deuterons}
\begin{figure}
\begin{center}
  {\includegraphics*[width=5.in]{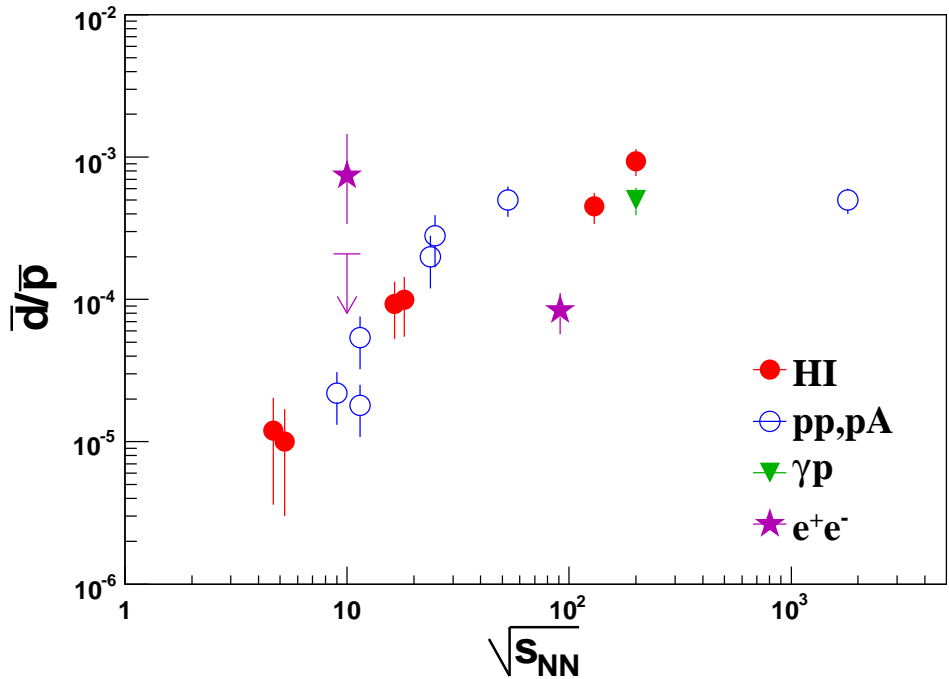}}
  \caption{ $\bar{d}/\bar{p}$ as a measure of antibaryon phase space
  density as a function of beam energy for various beam species.
  $e^+e^-$ and $\gamma p$ collisions are also shown at their center of
  mass beam energy.  }
  \label{phasespace_sqrts}
\end{center}
\end{figure}

\begin{table}[htb]
\begin{center}
\begin{tabular}{|c|c|c|} \hline
system & processes & $\bar{d}/\bar{p}$ \\
\hline
$e^+e^- (\Upsilon)$ & $ggg$ &$7.4^{+3.6}_{-2.0}10^{-4}$\\
$\gamma p (200)$   & $q\bar{q}g$&$5.0\pm1.1\times10^{-4}$\\
$pp (53)$  & $qg,\bar{q}g$ & $5.0\pm1.2\times10^{-4}$ \\
$p\bar{p} (1800)$  & $qg,\bar{q}g$ & $5.0\times10^{-4}$ \\
$AA (130)$  & $qg,\bar{q}g$ & $4.5\pm1.1\times10^{-4}$ \\
\hline
$e^+e^-(10)$ & $q\bar{q}$ &$<2.1\times10^{-4}$\\
$e^+e^-(91)$ & $q\bar{q}$ &$(8.4\pm2.7)\times10^{-5}$\\
$pp(A) (<20)$  & $qg,qq$ & $<10^{-4}$ \\
$AA (<20)$  & $qg,qq$ & $10^{-4}-10^{-5}$\\
\hline
\end{tabular}
\end{center}
\caption{Dominant processes in different collision systems and the
corresponding $\bar{d}/\bar{p}$ ratio. Values in the parentheses are
the center of mass beam energy in GeV.} \label{tab:qbarg}
\end{table}
We tabulate the collision system, their dominant processes and
$\bar{d}/\bar{p}$ in Tab.~\ref{tab:qbarg}.  Table~\ref{tab:qbarg} and
Fig.~\ref{phasespace_sqrts} show that collisions dominated by
$\bar{q}+g$ and/or $g+g$ saturate anti-baryon density at $10^{-3}$
while those dominated by $q+q(\bar{q})$ or $q+g$ produce much less
anti-baryons~\cite{argus,H1,aleph}. It is very convincing from the
collisions in $\gamma p$ and $e^+e^-$ at various energies.  The
measurements at $e^{+}e^{-}$ and $\gamma p$ may be used to gauge what
kind of partonic configuration creates baryons. There are two
measurements of $\bar{d}/\bar{p}$ from $e^{+}e^{-}$ at low
energies~\cite{argus}: one at $\Upsilon$ mass of $\sqrt{s}=9.86$ where
the final state hadrons are predominantly from $\Upsilon$ decay to
three gluons (${\it ggg}$) and the other an upper limit at continuous
energy of $\sqrt{s}=10.GeV$ where the final state hadrons come from
$q\bar{q}$ pair from a virtual photon. These two ratios are different
by more than a factor of 3. This may be related to how baryons are
created (more baryon from gluons than from quarks). In fact, not only
the $\bar{d}/\bar{p}$ are different in these two $e^{+}e^{-}$
collisions, the baryon production is higher at $\Upsilon$ than at the
continuous energy while the meson production is the same.  The
anti-baryon phase space density from ${\it ggg}$ configuration in
$e^{+}e^{-}$ is very similar to the anti-baryon phase space density
measured at RHIC. The $\bar{d}/\bar{p}$ ratio from the $\bar{q}q$
configuration is much lower. On the other hand, $\gamma p$
collisions~\cite{H1}, where the dominant process is $q\bar{q}+g$,
yield similar anti-baryon density as those in nucleus-nucleus
collisions.  

In addition, ALEPH Collaboration~\cite{aleph} found that baryon
production is suppressed in $e^+e^-$ to hadron event relative to other
system as shown in Fig.~\ref{phasespace_sqrts}. The double ratio of
$\bar{d}/\bar{p}$ in $e^+e^-$ between Z boson hadron decay events
($\bar{d}/\bar{p}=8.4\pm2.7\times10^{-5}$) and $\Upsilon\rightarrow
ggg$ events ($\bar{d}/\bar{p}=7.4{}^{+3.6}_{-2.0}\times10^{-4}$) is
$\gamma_{B}=0.11^{+0.05}_{-0.06}$.  In those Z boson hadron events,
the dominant process is $q\bar{q}$ fragmentation. However, the event
rate of process $q\bar{q}g$, whose fragmentation would yield same
$\bar{d}/\bar{p}$ as in $ggg$, is suppressed by a factor of strong
coupling constant $\alpha_{s}(M_{Z})=0.116$. We have shown in
Table~\ref{tab:qbarg} and Fig.~\ref{phasespace_sqrts} that production
of anti-deuterons is negligible in $q\bar{q}$ compared to $q\bar{q}g$
events, and both $ggg$ ($\Upsilon$) and $q\bar{q}g$ ($\gamma p$)
saturate $\bar{d}/\bar{p}$. The suppression of $\bar{d}/\bar{p}$ by a
factor of $\gamma_{B}=0.11$ in $e^+e^-$ at LEP can be readily
explained by the event rate $e^+e^-\rightarrow q\bar{q}g$ at
$\alpha_{s}(M_Z)=0.116$. This observation can be further tested by
measuring $\bar{d}/\bar{p}$ in $e^+e^-\rightarrow\bar{q}q$ and
$e^+e^-\rightarrow\bar{q}qg$ separately at LEP or other
facilities. The baryon production in a gluon jet is indeed about a
factor of 2 higher than that in a quark jet as measured separately at
LEP.  This will not explain a factor of 10 differences in baryon phase
space density.  On the other hand, the particle spectra in a gluon jet
are in general softer than those in a quark jet. This condensation in
momentum and coordinate spaces increases the phase space density of
the baryons.  From Table~\ref{tab:qbarg}, it is inconclusive whether
$q+g$ produces less anti-baryons at low beam energy due to low
production of anti-deuteron in this configuration or due to energy
threshold in producing anti-deuterons. In any case, it is conclusive
that baryon density from collisions involving a gluon is much higher
than those without gluon.
\section{Implications to Baryon/Meson Enhancement and Jet Quenching at RHIC}
Although what presented here are measurements of anti-deuteron
production at relatively low $p_T$, it requires a total energy of at
least $W=3.8$ GeV in an elementary collision.  This is not a soft
process even though coalescence is a final state interaction. In what
follows, we attempt to apply the observation that baryons
(anti-baryons) are produced by processes of $q+g$ and $g+g$
($\bar{q}+g$ and $g+g$) to the baryon enhancement at high $p_T$ at
RHIC when jet is quenched in the strongly interacting QCD matter.

At RHIC, baryons are enhanced relative to mesons toward more central
Au+Au collisions at intermediate
$p_T$~\cite{phenixstarPID,starphikstar,starv2} while elliptic flow  
is found to follow number of constituent quark scaling in the same
$p_T$ range~\cite{starv2}. Coalescence of quarks at hadronization  
can quantitatively explain these phenomena. At higher $p_T$, baryons  
are expected to be more suppressed than mesons since gluons, which  
fragment relatively more to baryons than mesons, are more suppressed  
than quarks due to a casimir color factor of 9/4 in parton energy  
loss when they traverse hot and dense QCD medium. Experimentally,  
the suppressions of baryons and mesons at high $p_T$ are observed to  
be the same by STAR at RHIC~\cite{AuAupid}. Compton-like scattering  
with leading quark and gluon conversions in the quark-gluon plasma  
can explain the phenomenon when large scattering cross section is  
assumed~\cite{wliu}.  On the other hand, if $q+g$ configuration can  
always enhance baryon production regardless of whether gluon or  
quark is leading parton, the required large parton cross section can  
be compromised.
\section{Conclusions}
In summary, we presented systematically the anti-baryon density
inferred from $\bar{d}/\bar{p}$ measurements in various collision
systems at various energies.  It was found that the density in $\gamma
p$, pp, pA and AA collisions follows a universal distribution as
function of beam energy and can be described statistically. The
anti-baryon density at the coalescence saturates when the processes
from different collisions involving gluons. The suppression of baryon
density at LEP is due to low event rate involving gluon jets.  We
stipulated a possible explanation of large $\bar{p}/\pi$ ratio in the
intermediate $p_T$ and high $p_T$ at RHIC.
\section{Acknowledgements}
The authors thank Drs. J.M. Engelage, E. Finch, N. George, D. Hardtke,
C.M. Ko, R. Majka, R. Rapp, F.Q. Wang and N. Xu for valuable
discussions. This work was supported in part by the HENP Divisions of
the Office of Science of the U.S. DOE; the Ministry of Education and
the NNSFC of China. XZB was supported in part by a DOE Early Career
Award and the Presidential Early Career Award for Scientists and
Engineers.


\end{document}